\documentclass[twocolumn]{article}
\usepackage{color}
\usepackage{amsmath}
\usepackage{amsfonts}
\usepackage{graphicx}

\definecolor{color1}{rgb}{0.000,0.000,0.000}
\definecolor{color2}{rgb}{0.000,0.000,1.000}
\definecolor{color3}{rgb}{0.000,1.000,1.000}
\definecolor{color4}{rgb}{0.000,1.000,0.000}
\definecolor{color5}{rgb}{1.000,0.000,1.000}
\definecolor{color6}{rgb}{1.000,0.000,0.000}
\definecolor{color7}{rgb}{1.000,1.000,0.000}
\definecolor{color8}{rgb}{1.000,1.000,1.000}
\definecolor{color9}{rgb}{0.000,0.000,0.502}
\definecolor{color10}{rgb}{0.000,0.502,0.502}
\definecolor{color11}{rgb}{0.000,0.502,0.000}
\definecolor{color12}{rgb}{0.502,0.000,0.502}
\definecolor{color13}{rgb}{0.502,0.000,0.000}
\definecolor{color14}{rgb}{0.502,0.502,0.000}
\definecolor{color15}{rgb}{0.502,0.502,0.502}
\definecolor{color16}{rgb}{0.753,0.753,0.753}

\begin{document}

\title{Embedding the Schwarzschild Ideal Fluid Metric}
\author{Earnest Harrison\footnote{email: \tt earnest.r.harrison@comcast.net}}
\maketitle

\begin{abstract}
Certain semi-Riemannian metrics can be decomposed into a Riemannian 
part and an isochronal part. The properties of such metrics are 
particularly easy to visualize in a coordinate-free way, using 
isometric embedding. We present such an isochronal, isometric embedding of the well known 
Schwarzschild ideal fluid metric in an attempt to see what is happening
when the pressure becomes singular. 
\end{abstract}

\paragraph{Introduction:}
We desire to visualize a manifold that admits to an ideal fluid metric and see how the pressure singularities appear in that manifold. While there are a large number of static, spherically symmetric, ideal fluid metrics in the literature (see, for example \cite{Stephani2} and \cite{Boonserm1}), we will choose the Schwarzschild ideal fluid metric.  Other ideal fluid metrics, with more realistic equations of state, have an increasing density towards the center of the sphere, which causes pressure singularities to appear at lower mass to radius ratios. Our goal is to push the mass to radius ratio as high as possible. 

To create an isometric embedding of a metric means to find a manifold that admits to the specified metric, and to place that manifold in some higher-dimensional flat space. The geodesics of this manifold will be exactly the same as those of the metric tensor, over the range of validity of the embedding.

We will be embedding in an $n+1$ Minkowski space, and will require $n+1$ functions, $y^i$, of the independent variables of the metric that satisfy the following equation \cite{Dirac}:
\begin{equation}
g_{\alpha \beta}=h_{ij} y^{i}_{,\alpha} y^{j}_{,\beta}
\label{eq:embedding1}
\end{equation}
where repeated Roman indices indicate summation from 0 to n, Greek indices range from $0$ to $3$, comma means ordinary partial differentiation, and $h_{ij}=\mathrm{Diagonal} \left[-1, 1, \dots 1 \right]$. Constraining $y^0$ to be equal to $c \, t$ makes the embedding isochronal.

\paragraph{The Metric and its Isochronal Embedding:} Schwarzschild's ideal fluid metric defines the squared length of a line segment for a fluid sphere of constant density. It is given by \cite{Stephani1}:
\begin{equation}
\begin{aligned}
ds^{2} =&\frac{dr^{2} }{1-A\,r^{2} } +r^{2} d\Theta ^{2} \\
&-\left(
\frac{3}{2} \sqrt{1-A\,r_{0}^{2} } -\frac{1}{2} \sqrt{1-A\,r^{2} } \right)
^{2} c^{2} dt^{2} 
\end{aligned}
\end{equation}
where
\begin{equation}
\begin{aligned}
d \Theta ^{2} &=d\theta ^{2} +\cos ^{2} \left( \theta \right) d\phi ^{2} \\
A&=\frac{2m}{r_{0}^{3} } ,\quad r\leq r_{0} ,\quad r_{0} \geq 2m
\end{aligned}
\end{equation}
and where $2m$ is the usual Schwarzschild radius of the mass.

Although we can work in Schwarzschild's original coordinates, isotropic coordinates have a slightly larger range of validity. The manifolds described by each approach are identical over the mutual range of validity.

The required coordinate change is one that will take us to this form of the metric:
\begin{equation}
\begin{aligned}
ds^{2} =g^{2} \left( dR^{2} +R^{2} d\Theta ^{2} \right) -f^{2} c^{2}
dt^{2}
\end{aligned}
\end{equation}
The transformation, which can be found from a first-order differential equation, is given by:
\begin{equation}
\begin{aligned}
r=&\frac{\left( 1+\rho _{0}^{2} \right) ^{3} }{1+\rho ^{2} } R \\
\textrm{where }
\rho^{2} =&\frac{m}{2R_{0} } \left( \frac{R}{R_{0} } \right) ^{2} ,\quad \rho
_{0}^{2} =\frac{m}{2R_{0} }
\end{aligned}
\end{equation}
where $R_{0}$ is the surface of the sphere in the new coordinate system.

Then the functions, $f(R)$ and $g(R)$, are found to be:
\begin{equation}
\begin{aligned}
f =\frac{3}{2} \,\,\frac{1-\rho _{0}^{2} }{1+\rho _{0}^{2} } -\frac{1}{2}
\,\,\frac{1-\rho ^{2} }{1+\rho ^{2} },\quad 
g =\frac{\left( 1+\rho _{0}^{2} \right) ^{\,3} }{1+\rho ^{2} } 
\end{aligned}
\label{eq:rho}
\end{equation}
$\rho =\rho_{0}$ corresponds to the surface of the fluid sphere. 

The ideal fluid metric and the vacuum metric must have the same value at the surface of the sphere. By Birkoff's theorem, the vacuum metric can be found by replacing $\rho_{0}$ with $\rho$ in Equation \ref{eq:rho}. Making this substitution forms a $C^{1}$ interface to the ideal fluid solution at $R = R_{0}$:
\begin{equation}
\begin{aligned}
f_{v} =\frac{1-\rho _{v}^{2} }{1+\rho _{v}^{2} },\quad 
g_{v} =\left( 1+\rho _{v}^{2} \right) ^{2},\quad 
\rho _{v}^{2} =\frac{m}{2R}  
\end{aligned}
\label{eq:vacuum}
\end{equation}
where the subscript, $v$, is intended to denote the vacuum region. 

Focusing on the fluid region and using the methods of \cite{Harrison} 
we arrive at the embedding functions:
\begin{equation}
\begin{aligned}
y^{0}  & = c\thinspace t \\
y^{1}  & = 2k\thinspace m\thinspace \sqrt{1-f^{2} } \cos \left(
\frac{c\thinspace t}{2k\thinspace m} \right)  \\
y^{2}  & = 2k\thinspace m\thinspace \sqrt{1-f^{2} } \sin \left(
\frac{c\thinspace t}{2k\thinspace m} \right)  \\
y^{3}  & = g\thinspace R\cos \left( \theta \right) \cos \left( \phi
\right)  \\
y^{4}  & = g\thinspace R\cos \left( \theta \right) \sin \left( \phi
\right)  \\
y^{5}  & = g\thinspace R\sin \left( \theta \right)  \\
y^{6}  & = \int \sqrt{\, g^{2} -\left( g+R \,
g' \right)^{2} - \frac{\left( 2 \, k \, m \,f \, f' \right)^2}{1 - f^{2}}} \ dR
\end{aligned}
\label{eq:embedding2}
\end{equation}
Where $k$ is a constant.

\begin{figure}[htbp]
\includegraphics[width=0.49\textwidth]{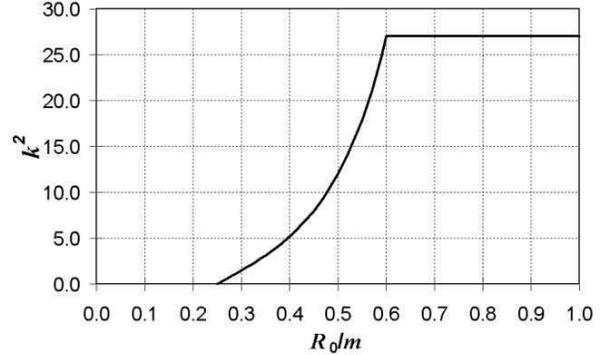}
\caption{Limits on $k^{2}$ as a function of $R_{0}$ so that the embedding remains real.}
\label{fig:k-limit}
\end{figure}

This embedding may be thought of as a Riemannian manifold spinning in Minkowski space at an angular rate of $\omega = c/(2k \thinspace m)$ radians per second, as long as all of the embedding functions are real. This will be true for each of the equations in \ref{eq:embedding2} when $R\geq 0$ and $k^{2} $ is below the curve in Figure \ref{fig:k-limit}. Had we remained in the original coordinate system of Equation 1, we would have been limited to $R_{0} \geq m/2 $ (or $r_{0} \geq 2 m$ in the original coordinate system).

Once we have selected $R_{0}$ and $k$, the manifold behaves as it were a physical object, and we have a coordinate-free tool for studying the underlying metric. While we had to use a metric and its coordinate system to find the manifold, the manifold itself does not require a coordinate system to exist. A coordinate system is simply a scheme to name the points in the manifold.

\begin{figure}[htbp]
\includegraphics[width=0.49\textwidth]{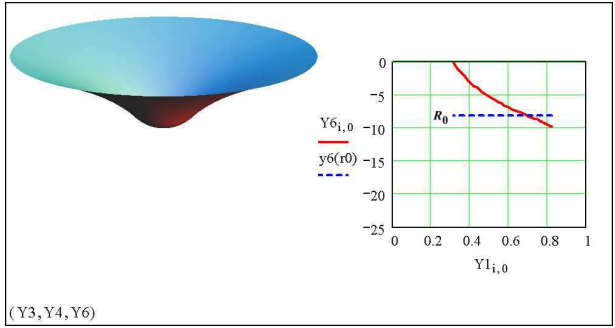}
\caption{Manifold slices $y^3 - y^4 - y^6$, and $y^1 - y^6$ at $t = \phi=0$ with $R_{0} = 3 m$, marked with the dashed line. The manifold rotates in the $y^1 - y^2$ plane.}
\label{fig:3000}
\end{figure}

\begin{figure}[htbp]
\includegraphics[width=0.49\textwidth]{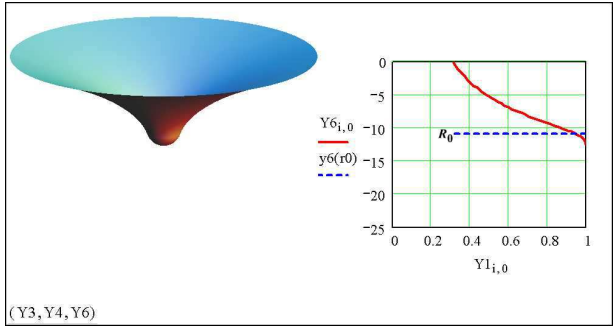}
\caption{Setting $R_{0} = m$ causes a pressure singularity to appear at $R=0$,
where the rotation speed equals $c$.}
\label{fig:1000}
\end{figure}

\begin{figure}[htbp]
\includegraphics[width=0.49\textwidth]{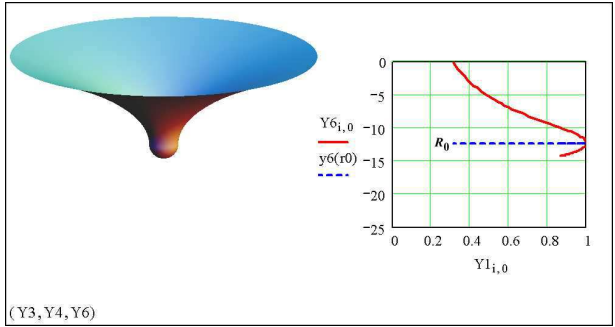}
\caption{Setting $R_{0} = 0.5 m$ places the event horizon at the surface of the sphere.}
\label{fig:0500}
\end{figure}

\begin{figure}[htbp]
\includegraphics[width=0.49\textwidth]{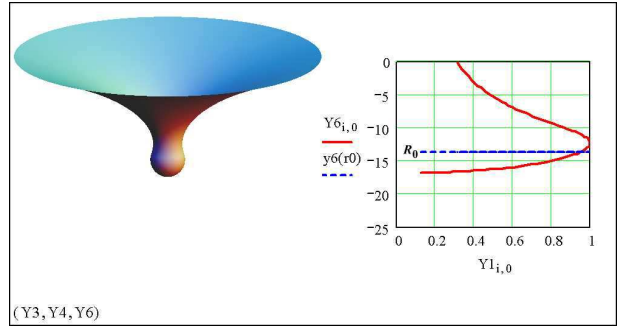}
\caption{At $R_{0} = 0.251 m$, the event horizon remains at $m/2$, and the pressure, while negative, is finite.}
\label{fig:0250}
\end{figure}

\paragraph{Visualization:}
If a manifold is like a physical object, we should be able to draw a picture of it, or at least slices of it. Figures \ref{fig:3000} through \ref{fig:0250} show slices of this manifold for various values of $R_{0} /m $. They depict a slice through the manifold at $t=0$ and $\phi = 0$ such that $y^0, y^2$ and $y^5$ are zero. 

Since we cannot depict the remaining four dimensions in one figure, we have broken the manifold into the $y^3, y^4$ and $y^6$ cut and a $y^1, y^6$ cut which we have placed beside the first. The scale has been adjusted so that the two vertical axes approximately correspond. Each point on the curve on the right should be thought of as representing a sphere whose diameter is that of the circle formed by a cut through the left figure at the height of that point.

The curve on the right of each figure pair is spinning in the $y^1 - y^2$ plane. Thus the whole manifold is spinning in this plane within the Minkowski space. The horizontal axis is calibrated such that the speed of the manifold, $c \thinspace \sqrt{1-f^{2}}$, is relative to the speed of light.

There are no boundaries in these manifolds. The top edge is where we elected to stop plotting. The bottom end of the line in the $y^1, y^6$ cut, corresponds to the center of the sphere and thus is not a boundary.

The pressure becomes singular at the center once $R_{0}=m$. A sphere of singular pressure forms if $R_{0} < m$. This pressure singularity occurs at the radius where the velocity reaches unity. There is nothing else particularly notable about this radius; the manifold is smooth there. Negative pressure can be seen to be simply the force of the fluid below the critical radius being driven upward to this particular radius.

\begin{figure}[htbp]
\includegraphics[width=0.49\textwidth]{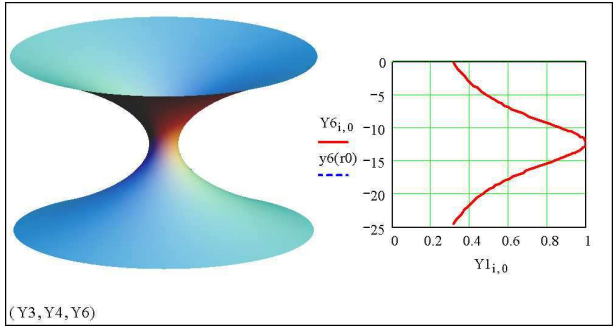}
\caption{The vacuum, black hole condition corresponds to $R_{0} = 0$. The fluid-vacuum interface has moved off the bottom of the diagram.}
\label{fig:0000}
\end{figure}

We can take $R_{0}$ to zero and produce a valid embedding as shown in Figure \ref{fig:0000}. This manifold corresponds to a vacuum solution of Einstein's equation.\footnote{This manifold differs from the one depicted in \cite{Harrison} in that here, $k$ has been set to a small value to be consistent with the other figures.}
One can also produce plots with $R_{0} < 0.25 m$ and see a smooth progression from Figure \ref{fig:0250} to Figure \ref{fig:0000}, but the embeddings are only valid for a piece of the manifolds; $y^{2}$ (at least) ceases to be real near the origin when $0 < R_{0} < 0.25 m$.

\paragraph{Other Embeddings:}
Embeddings are not unique. In addition to rotations (including Lorentz transformations), translations, reflections, and changes to parameters like $k$, we can make different selections of the embedding functions to satisfy Equation \ref{eq:embedding1}. Kruskal \cite{Kruskal} and Fronsdal \cite{Fronsdal} independently produced an embedding different than the one presented here. 

Kruskal's maximal extension of the Schwarzschild metric  may be viewed as a a partial embedding; he confined his attention to the $r-t$ subspace of that metric. Fronsdal published an embedding of the Schwarzschild vacuum metric that is fully consistent with Kruskal's work, but includes the other dimensions. His embedding is given by:
\begin{equation}
\begin{aligned}
y^{0} &=\left\{ 
\begin{array}{cc}
4m\sqrt{1-\frac{2m}{r} } \sinh \left( \frac{ct}{4m} \right)  & r\geq 2m \\
4m\sqrt{\frac{2m}{r}-1 } \cosh \left( \frac{ct}{4m} \right)  & r < 2m \\
\end{array}
\right.   \\
y^{1} &=\left\{ 
\begin{array}{cc}
4m\sqrt{1-\frac{2m}{r} } \cosh \left( \frac{ct}{4m} \right)  & r\geq 2m \\
4m\sqrt{\frac{2m}{r}-1 } \sinh \left( \frac{ct}{4m} \right)  & r < 2m \\
\end{array}
\right.  \\
y^{2} &=r\cos \left( \theta \right) \cos \left( \phi \right)   \\
y^{3} &=r\cos \left( \theta \right) \sin \left( \phi \right)   \\
y^{4} &=r\sin \left( \theta \right)   \\
\end{aligned}
\label{eq:fronsdal}
\end{equation}
and:
\begin{equation*}
y^{5} =\int \sqrt{\frac{2m}{r} +\left( \frac{2m}{r} \right) ^{2} +\left(
\frac{2m}{r} \right) ^{3} } dr  
\end{equation*}
where $y^{0}$ is time like.

Fronsdal's embedding defines a manifold with a boundary. Fronsdal argued that, by adding a second manifold with  $y^{0}$ and $y^{1}$ replaced by $-y^{0}$ and $-y^{1}$, respectively, then letting $t$ pass to infinity and gluing the edges together, one can eliminate the boundary. The resulting construction is essentially identical to Kruskal's maximal extension, except Kruskal used a singular transformation of coordinates to avoid the taking of limits.

\begin{figure}[htbp]
\includegraphics[width=0.49\textwidth]{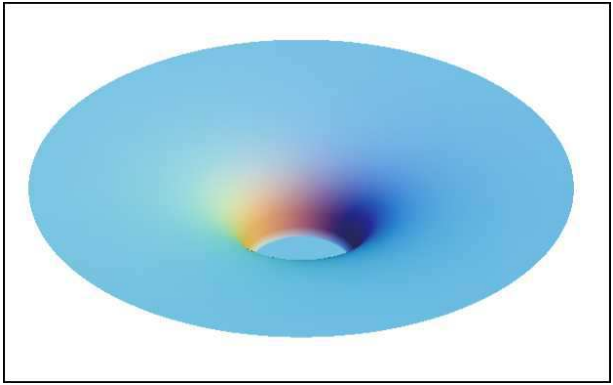}
\caption{Plotting the $y^{2}, y^{3}, y^{1}$ cut shows the manifold has a corner when $t=0$.}
\label{fig:frons1}
\end{figure}

\begin{figure}[htbp]
\includegraphics[width=0.49\textwidth]{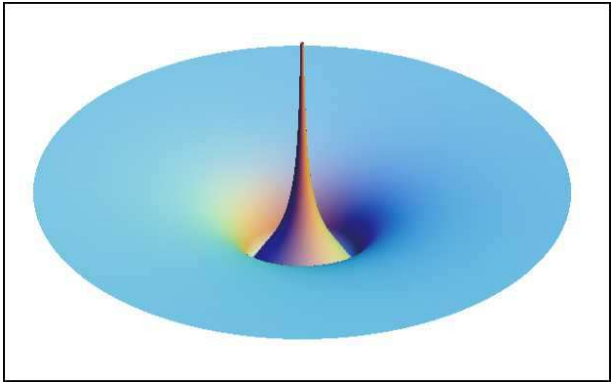}
\caption{Setting $t$ slightly positive, reveals a cusp at the event horizon and a singularity at the origin.}
\label{fig:frons2}
\end{figure}

\begin{figure}[htbp]
\includegraphics[width=0.49\textwidth]{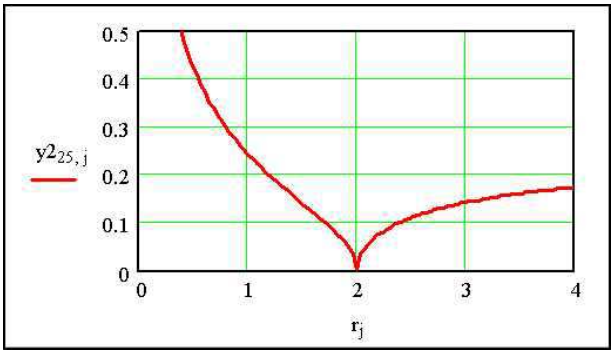}
\caption{A radial cut through the section in Figure \ref{fig:frons2}, clearly shows the cusp.}
\label{fig:cusp}
\end{figure}

Unfortunately, there are difficulties with Fronsdal's embedding. Let us examine the $y^{2}, y^{3}, y^{1}$ ($t=\phi=0$) cut through the space. As seen in Figure \ref{fig:frons1}, there is a corner in the ``manifold'' (more properly now, a ``variety''). More troubling is when $t>0$, an example of which is depicted in Figures \ref{fig:frons2} and \ref{fig:cusp}. We now see that the corner has turned into a cusp and the center has become singular.

Looking closer, we see that the first derivatives, $y^{0}_{,r}$ and $y^{1}_{,r}$, are singular at both $r=2m$ and $r=0$. These are not simple coordinate singularities. They are present in the variety itself and cannot be removed through any coordinate transformation. The manifold has these properties regardless of how we name the points in that manifold.


While it is undeniable that Einstein's vacuum equation is satisfied on either side of the event horizon and any point away from the origin, one must question if there is any meaning to differential equations applied to the event horizon or origin.

We believe that the confusion arises because Einstein's equations are second order, and thus should admit to two solutions. And it turns out that we have two choices as to how to proceed across the event horizon: Fronsdal's embedding or the isochronal embedding. Fronsdal's embedding has not satisfied anything more than $C^{0}$ continuity at this boundary. The isochronal embeddings presented here have $C^{\infty}$ continuity everywhere, except at the fluid-vacuum surface, where the continuity is $C^{1}$. (This reduction in continuity is caused by the specified step-discontinuity in density at this interface.) We need at least $C^{1}$ for Einstein's second-order partial differential equations to have any meaning.

\paragraph{Conclusion:}
The manifolds that we have displayed correspond to static metrics. We have pushed the ideal fluid solution well beyond the point of a realizable physical object; no matter could ever hold back the pressure. However, when a star first starts to collapse, the manifold describing that collapse would likely have a shape similar to these isochronal manifolds. Thus there may be a singularity-free solution to the stellar collapse event.


\begin{thebibliography}{9}

\bibitem{Stephani2} Stephani, H., \textit{et. al., Exact Solutions of Einstein's Field Equations}, Cambridge University Press, (New York: 2003).
\bibitem{Boonserm1} Boonserm, P, \textit{et. al.}, ``Generating Perfect Fluid Spheres in General Relativity'', arXiv:gr-qc/0503007 (March, 2005).
\bibitem{Dirac} Dirac, P.A.M., \textit{General Theory of Relativity}, John Wiley \& Sons, (New York: 1975), p 11.
\bibitem{Stephani1}  Stephani, H., \textit{General Relativity, An Introduction to the Theory of the Gravitational Field}, Cambridge University Press, (New York: 1990).
\bibitem{Harrison} Harrison, E.R., ``An Explicit Isochronal, Isotropic Embedding'', arXiv:gr-qc/0601027 (January, 2006).

\bibitem{Kruskal} Kruskal, M.D. ``Maximal extension of Schwarzschild Metric'', Physical Review, 119 (1960) p 1743.
\bibitem{Fronsdal} Fronsdal, C., ``Completion and Embedding of the Schwarzschild Solution'', Phys Rev No. 116, p 778 (1959).


\end{thebibliography}
\end{document}